\documentclass[submission, Phys]{SciPost}
\usepackage{epstopdf}
\usepackage{multirow}
\begin{document}

\begin{center}{\Large \textbf{Linear optical universal quantum gates with higher success probabilities
}}\end{center}

\begin{center}
Wen-Qiang Liu\textsuperscript{1}  and
Hai-Rui Wei\textsuperscript{2,*}
\end{center}

\begin{center}
{\bf 1} Center for Quantum Technology Research and Key Laboratory of Advanced Optoelectronic Quantum Architecture and Measurements (MOE), School of Physics, Beijing Institute of Technology, Beijing 100081, China
\\
{\bf 2} School of Mathematics and Physics, University of Science and Technology Beijing, Beijing 100083, China
\\
* hrwei@ustb.edu.cn
\end{center}

\begin{center}
\today
\end{center}


\section*{Abstract}
{\bf
Universal quantum gates lie at the heart of designing quantum computer.  We construct two compact quantum circuits to implement post-selected controlled-phase-flip (CPF) gate and Toffoli gate with linear optics assisted by one and two single photons, respectively. The current existing maximum success probability of 1/4 for linear optical CPF gate is achieved by resorting to an ancillary single photon rather than an entangled photon pair or two single photons. Remarkably, our Toffoli gate is accomplished with current maximum success probability of 1/30 without using additional entangled photon pairs and the standard decomposition-based approach. Linear optical implementations of the presented two universal gates are feasible under current technology and provide a potential application in large-scale optical quantum computing.
}

\vspace{10pt}
\noindent\rule{\textwidth}{1pt}
\tableofcontents\thispagestyle{fancy}
\noindent\rule{\textwidth}{1pt}
\vspace{10pt}

\section{Introduction}\label{sec1}


Quantum computing  \cite{book} has the remarkable potential to dramatically surpass its classical counterpart on solving certain complex tasks in terms of the
processing speed or resource overhead. Universal quantum gates are crucial building blocks in quantum circuit model \cite{Barenco,circuits,Fredkin-Liu1,Fredkin-Liu2,Nikolaeva}, quantum algorithms \cite{Grover,Shor,Bharti}, quantum simulations \cite{Georgescu,wang2018efficient,he2021quantum}, quantum communication \cite{Pan2}, and quantum computing \cite{gui2006general,zou2009mathematical,wei2016duality,song2021heralded}. Photon is generally viewed as one of the promising candidates for flying and solid-state quantum computing owing to its outstanding low decoherence, high-speed transmission, natural information carrier, flexible single-qubit manipulations, and available atom-like qubit interconnector \cite{computing0,computing}. Strong interactions between individual photons are the key resources for nontrivial multi-photon quantum gates, and the prohibited photon-photon interactions can be remedied efficiently by using linear optics \cite{KLM} or solid-state media \cite{QD,NV,atom}. Unfortunately, solid-state platforms are challenged  by inefficiency and imperfection in experiments. The probabilistic character of universal quantum gates with linear optics is unavoidable. Hence, minimizing the quantum resources required to implement quantum gates with higher success probability is a central problem of linear optical quantum computing, and tremendous efforts have been made on it \cite{CNOT-KLM,CNOT-Bell-1/4-1,CNOT-Bell-1/4-2,CNOT-Bell-1/4-3,CNOT-Bell-1/4-4,CNOT-Bell-1/4-5,CNOT-1/9-2,CNOT-1/9-4,CNOT-1/9-5,CNOT-1/9-3,CNOT-1/9-6,CNOT-1/9-7,CNOT-1/8-1,CNOT-1/8-2,CNOT-1/16,CNOT-SR}.

Controlled phase flip (CPF) gate or its equivalent controlled-NOT (CNOT) gate is the most quintessential universal quantum gate \cite{book}. CNOT gates together with single-qubit rotations are sufficient to implement any quantum computation \cite{Barenco}. Nowadays, CNOT gate has been experimentally demonstrated in several physical systems \cite{CNOT-superconducting,CNOT-atom,CNOT-ion}.
The KLM scheme \cite{KLM} is served as a stepping stone for implementing CPF gate with a sheer number of linear optics, large and good quantum memory, and giant interferometer phase stable. Various improved works were later proposed  both in theory and experiment \cite{CNOT-KLM,CNOT-Bell-1/4-1,CNOT-Bell-1/4-2,CNOT-Bell-1/4-3,CNOT-Bell-1/4-4,CNOT-Bell-1/4-5,CNOT-1/9-2,CNOT-1/9-4,CNOT-1/9-5,CNOT-1/9-3,CNOT-1/9-6,CNOT-1/9-7,CNOT-1/8-1,CNOT-1/8-2,CNOT-1/16,CNOT-SR}.
So far, it has been demonstrated that CPF gate can be completed with a success probability of 1/9, which is the existing maximum value achievable without ancillary photons \cite{CNOT-1/9-2,CNOT-1/9-4,CNOT-1/9-5,CNOT-1/9-3,CNOT-1/9-6,CNOT-1/9-7},
and the success probability can be improved to 1/8 via two additional independent single photons \cite{CNOT-1/8-1,CNOT-1/8-2}.
The current existing highest success probability of 1/4 for a CNOT gate has been achieved assisted by a necessary entangled photon pair  \cite{CNOT-Bell-1/4-1,CNOT-Bell-1/4-2,CNOT-Bell-1/4-3,CNOT-Bell-1/4-4,CNOT-Bell-1/4-5}. Deterministic generation of entangled photon pairs based on spontaneous parametric down-conversion remains a key technical obstacle  in experiments due to multi-photon emissions and probabilistic  properties \cite{SPDC}.

Toffoli gate supplemented with Hadamard gate  can simulate any multi-qubit quantum computing \cite{book}.  Toffoli gate is also served as an essential part in  quantum factoring algorithm \cite{Shor1}, quantum search algorithm \cite{Grover1}, quantum half-adder \cite{adder}, quantum error correction \cite{correction1}, and quantum fault tolerance \cite{tolerant1}, etc. Much attention has been paid to the realization of Toffoli gate \cite{Toffoli-ion,Toffoli-superconducting,Toffoli-atom}.
It has been confirmed theoretically that the optimal cost of a Toffoli gate is six CNOT gates \cite{Barenco} or five two-qubit entangling gates \cite{five}. Such synthesis might be helpful to design complex quantum gate, but it makes the gate further susceptible to the environmental noise  and increases the time scale of the system.
Without using the standard decomposition-based approach,  early in 2006, Fiur\'{a}\v{s}ek  \cite{T-Fiurasek} first showed a three-photon polarization Toffoli gate with a success probability of 0.75\% (approximately  1/133) using linear optics.  Using higher-dimensional Hilbert spaces, Ralph \emph{et al.} \cite{T-PRA} improved the success probability of  a linear optical Toffoli gate to 1/72 in 2007, and this interesting probabilistic  scheme was later experimentally demonstrated in 2009 \cite{T-NatPhy}. Recently, Liu \emph{et al.} \cite{T-Liu} further enhanced the success probability of the Toffoli gate to 1/64. In the same year, a post-selected Toffoli gate with a success probability of 1/60 was experimentally realized on a programmable
silicon-based photonic chip \cite{li2022quantum}.  Additionally, in 2022, Li \emph{et al.} \cite{T-path} experimentally demonstrated a path-based three-photon Toffoli gate with a success probability of 1/72.  Many hybrid multiple degrees of freedom (DOFs)  probabilistic and deterministic Toffoli gates were also reported in recent years \cite{hybrid1,hybrid2,hybrid3,hybrid4,hybrid5}.


In this paper, we propose two compact quantum circuits to implement post-selected CPF  and Toffoli gates in the coincidence basis using solely polarizing beam splitters (PBSs), half-wave plates (HWPs), beam splitters (BSs), and single-photon detectors.
Assisted by one and two independent single photons, our CPF and Toffoli gates are accomplished respectively when exactly one photon appears in each output mode.
Our schemes are appealing for higher success probabilities and less quantum resource requirements.
The existing highest success probability of a linear optical CPF gate 1/4 is achieved resorting to an auxiliary single photon in our scheme rather than an auxiliary entangled photon pair \cite{CNOT-Bell-1/4-1,CNOT-Bell-1/4-2,CNOT-Bell-1/4-3,CNOT-Bell-1/4-4,CNOT-Bell-1/4-5}.
The presented CPF gate also beats the ones with the success probability of 1/8 assisted by two single photons \cite{CNOT-1/8-1,CNOT-1/8-2}, and the ones with 1/9 without auxiliary photons \cite{CNOT-1/9-2,CNOT-1/9-4,CNOT-1/9-5,CNOT-1/9-3,CNOT-1/9-6,CNOT-1/9-7}.
In addition, the average success probability of our Toffoli gate is high to 1/30, which far exceeds all previous results for the same works \cite{T-Fiurasek,T-PRA,T-NatPhy,T-Liu}.



\section{Post-selected CPF gate with linear optics} \label{sec2}

It is well-known that CPF gate introduces a $\pi$ phase shift when the first qubit and the second qubit are both $|1\rangle$, and the rest remains unchanged. We encode the gate qubit in two polarization DOFs of a single photon, i.e., the horizontally polarized photon $|H\rangle=|0\rangle$ and vertically polarized photon $|V\rangle=|1\rangle$, respectively.

\begin{figure} [htpb]
\begin{center}
\includegraphics[width=9.5 cm,angle=0]{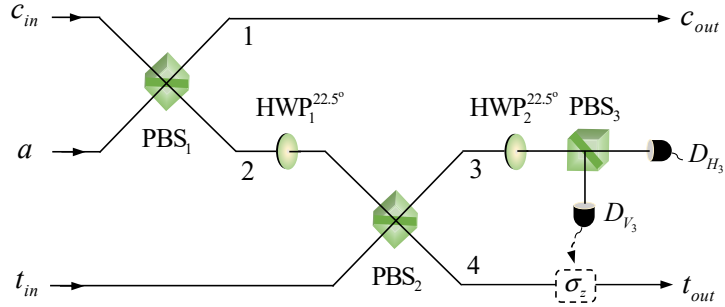}
\caption{ Schematic diagram of a post-selected CPF gate.  The gate is completed in case each spatial modes 1, 3 and 4 contain exactly one photon. PBS, a polarizing beam splitter, transmits the $H$-polarized photon and reflects the $V$-polarized photon. HWP$^{22.5^\circ}$, a half-wave plate setting at $22.5^\circ$, results in $|H\rangle\rightarrow\frac{1}{\sqrt{2}}(|H\rangle+|V\rangle)$ and
$|V\rangle\rightarrow\frac{1}{\sqrt{2}}(|H\rangle-|V\rangle)$. $D_{H_3}$ and $D_{V_3}$ stand for the single-photon detectors. Feed-forward operation $\sigma_z=|H\rangle \langle H| -|V\rangle \langle V|$ is applied when $D_{V_3}$ is triggered. } \label{CPF}
\end{center}
\end{figure}

Our scheme described in Fig. \ref{CPF} shows a polarization-based post-selected CPF gate can be completed in the following three steps.

First, the two gate photons and one auxiliary photon in the states
\begin{eqnarray}              \label{eq1}
\begin{split}
|\phi\rangle_{c_{in}}=\alpha_1|H\rangle_{c_{in}}+\beta_1|V\rangle_{c_{in}},
\end{split}
\end{eqnarray}
\begin{eqnarray}              \label{eq2}
\begin{split}
|\phi\rangle_{t_{in}}=\alpha_2|H\rangle_{t_{in}}+\beta_2|V\rangle_{t_{in}},
\end{split}
\end{eqnarray}
\begin{eqnarray}              \label{eq3}
\begin{split}
|\phi\rangle_a=\frac{1}{\sqrt{2}}(|H\rangle_a+|V\rangle_a),
\end{split}
\end{eqnarray}
are injected into the spatial modes  $c_{in}$, $t_{in}$, and $a$, respectively. Here coefficients $\alpha_1$, $\beta_1$, $\alpha_2$, $\beta_2$ satisfy the conditions  $|\alpha_1|^2+|\beta_1|^2=1$ and $|\alpha_2|^2+|\beta_2|^2=1$. The subscripts denote the spatial modes of photons (also named photon's paths).

The photons emitted from spatial modes $c_{in}$ and $a$ are fed into PBS$_1$, simultaneously.  PBS$_1$ transforms the state of the whole system from $|\Phi_1\rangle =|\phi\rangle_{c_{in}}\otimes|\phi\rangle_a\otimes|\phi\rangle_{t_{in}}$ to
\begin{eqnarray}              \label{eq4}
\begin{split}
|\Phi_2\rangle =&\frac{1}{\sqrt{2}} \big(\alpha_1|H\rangle_1|H\rangle_2+\beta_1|H\rangle_1|V\rangle_1+\alpha_1|H\rangle_2|V\rangle_2+\beta_1|V\rangle_1|V\rangle_2\big)\\
&\otimes\big(\alpha_2|H\rangle_{t_{in}}+\beta_2|V\rangle_{t_{in}}\big).
\end{split}
\end{eqnarray}
Based on Eq. (\ref{eq4}), one can see that PBS$_1$ can complete a  parity-check measurement on the polarization photons by choosing the instance in which each of the spatial mode contains exactly one photon in post-selection principle, and then the system would be changed into the state
\begin{eqnarray}              \label{eq5}
\begin{split}
|\Phi_3\rangle =\big(\alpha_1|H\rangle_1|H\rangle_2+\beta_1|V\rangle_1|V\rangle_2 \big)
\otimes\big(\alpha_2|H\rangle_{t_{in}}+\beta_2|V\rangle_{t_{in}} \big),
\end{split}
\end{eqnarray}
with a probability of 1/2. While the instance in which each spatial mode involves two photons or none photon indicates the gate operation fails.

Second, as shown in Fig. \ref{CPF}, before and after the photons from modes 2 and $t_{in}$  pass through PBS$_2$  simultaneously, two polarization Hadamard operations are performed on them by using HWP$_1^{22.5^\circ}$ and HWP$_2^{22.5^\circ}$, respectively. Here half-wave plate oriented at 22.5$^\circ$ (HWP$^{22.5^\circ}$) completes the transformations
\begin{eqnarray}                  \label{eq6}
\begin{split}
&|H\rangle\xrightarrow{\text{HWP}^{22.5^\circ}}\frac{1}{\sqrt{2}}\big(|H\rangle+|V\rangle\big), \quad
&|V\rangle\xrightarrow{\text{HWP}^{22.5^\circ}}\frac{1}{\sqrt{2}}\big(|H\rangle-|V\rangle\big).
\end{split}
\end{eqnarray}
Operations $\text{HWP}_1^{22.5^\circ} \rightarrow \text{PBS}_2 \rightarrow \text{HWP}_2^{22.5^\circ}$ transform the state $|\Phi_3\rangle$ into
\begin{eqnarray}              \label{eq7}
\begin{split}
|\Phi_4\rangle =&\frac{1}{2}\big( \alpha_1\alpha_2|H\rangle_1|H\rangle_4+\alpha_1\beta_2|H\rangle_1|V\rangle_4
+\beta_1\alpha_2|V\rangle_1|H\rangle_4-\beta_1\beta_2|V\rangle_1|V\rangle_4 \big)|H\rangle_3\\&
+\frac{1}{2}\big( \alpha_1\alpha_2|H\rangle_1|H\rangle_4-\alpha_1\beta_2|H\rangle_1|V\rangle_4
+\beta_1\alpha_2|V\rangle_1|H\rangle_4+\beta_1\beta_2|V\rangle_1|V\rangle_4\big)|V\rangle_3\\&
+\frac{1}{2\sqrt{2}} (\alpha_1\alpha_2|H\rangle_1-\beta_1\alpha_2|V\rangle_1)(|H\rangle_3+|V\rangle_3)(|H\rangle_3-|V\rangle_3)\\&
+\frac{1}{\sqrt{2}}\big( \alpha_1\beta_2|H\rangle_1+\beta_1\beta_2|V\rangle_1\big)|H\rangle_4|V\rangle_4.
\end{split}
\end{eqnarray}
We choose the case where exactly one photon in each of the spatial modes 3 and 4, and then the system would be in a normalization state
\begin{eqnarray}              \label{eq8}
\begin{split}
|\Phi_5\rangle =&\frac{1}{\sqrt{2}}\big( \alpha_1\alpha_2|H\rangle_1|H\rangle_4+\alpha_1\beta_2|H\rangle_1|V\rangle_4
+\beta_1\alpha_2|V\rangle_1|H\rangle_4-\beta_1\beta_2|V\rangle_1|V\rangle_4 \big)|H\rangle_3\\&
+\frac{1}{\sqrt{2}}\big( \alpha_1\alpha_2|H\rangle_1|H\rangle_4-\alpha_1\beta_2|H\rangle_1|V\rangle_4
+\beta_1\alpha_2|V\rangle_1|H\rangle_4+\beta_1\beta_2|V\rangle_1|V\rangle_4\big)|V\rangle_3,
\end{split}
\end{eqnarray}
with a probability of $\frac{1}{2}\times\frac{1}{2}=\frac{1}{4}$.

Finally, the photon emitted from spatial mode $a$ will be detected by using PBS$_3$ and photon detectors $D_{H_3}$ and $D_{V_3}$.
Based on Eq. (\ref{eq8}), one can see that when $D_{H_3}$ is fired, the photons emitted from $c_{out}$ and $t_{out}$ kept are in the state
\begin{eqnarray}              \label{eq9}
\begin{split}
|\Phi_6\rangle =\alpha_1\alpha_2|H\rangle_1|H\rangle_4+\alpha_1\beta_2|H\rangle_1|V\rangle_4
+\beta_1\alpha_2|V\rangle_1|H\rangle_4-\beta_1\beta_2|V\rangle_1|V\rangle_4,
\end{split}
\end{eqnarray}
with a probability of $\frac{1}{4}\times\frac{1}{2}$. And then, the performance of CPF gate is completed.

When $D_{V_3}$ is fired, the system will collapse into the state
\begin{eqnarray}              \label{eq10}
\begin{split}
|\Phi_6'\rangle =\alpha_1\alpha_2|H\rangle_1|H\rangle_4-\alpha_1\beta_2|H\rangle_1|V\rangle_4
+\beta_1\alpha_2|V\rangle_1|H\rangle_4+\beta_1\beta_2|V\rangle_1|V\rangle_4,
\end{split}
\end{eqnarray}
with a probability of $\frac{1}{4}\times\frac{1}{2}$. It is easily to convert Eq. (\ref{eq10}) to Eq. (\ref{eq9}) by applying a feed-forward $\sigma_z$ operation  on the photon emitted from spatial mode 4, which can be achieved by an $\text{HWP}^{0^\circ}$.

Putting all the pieces together, one can see that the quantum circuit shown in Fig. \ref{CPF} completed a CPF operation conditional on  exactly one photon in each of the output spatial modes. The total success probability of the presented gate can reach the current existing best result $\frac{1}{8}+\frac{1}{8}=\frac{1}{4}$, and only one additional single photon is required.

The scheme shown in Fig. \ref{CPF} is actually the post-selection result, i.e., the gate success is heralded by simultaneous successful detection of exactly one photon for each qubit. Nowadays, the various schemes for realizing post-selected-based CPF gate \cite{CNOT-1/9-2,CNOT-1/9-4,CNOT-1/9-5} and CNOT gate \cite{CNOT-1/9-3,CNOT-1/9-6,CNOT-1/9-7} have been proposed and experimentally demonstrated. Our scheme surpasses them in terms of the success probability and the quantum resource cost.

\section{Post-selected Toffoli gate with linear optics} \label{sec3}

\begin{figure*}    [htbp]
\begin{center}
\includegraphics[width=14.7 cm,angle=0]{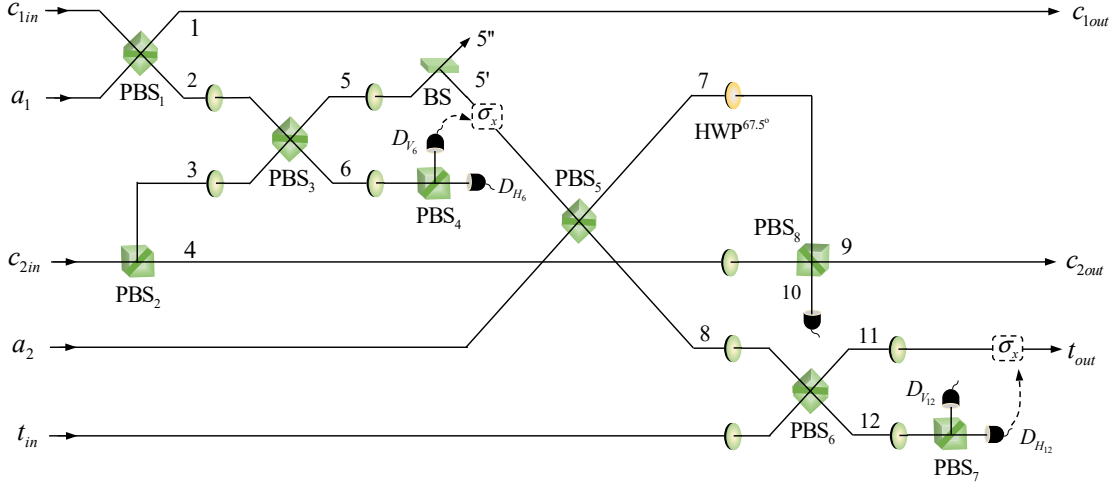}
\caption{Schematic diagram of a post-selected Toffoli gate. The gate works successfully conditioned on exactly one photon in each spatial modes 6, 12, $c_{1out}$, $c_{2out}$, and $t_{out}$. BS, a 50:50 beam splitter, results in $|H\rangle_5\rightarrow\frac{1}{\sqrt{2}}\big(|H\rangle_{5'}+|H\rangle_{5''}\big)$ and
$|V\rangle_5\rightarrow\frac{1}{\sqrt{2}}\big(|V\rangle_{5'}+|V\rangle_{5''}\big)$.  HWP$^{67.5^\circ}$ completes the transformations $|H\rangle\rightarrow\frac{1}{\sqrt{2}}(-|H\rangle+|V\rangle)$  and
$|V\rangle\rightarrow\frac{1}{\sqrt{2}}(|H\rangle+|V\rangle)$.  Pauli operation $\sigma_x=|H\rangle \langle V|+ |V\rangle \langle H|$ is applied when $D_{V_6}$ or
$D_{H_{12}}$ is triggered, which is achieved by an HWP$^{45^\circ}$.} \label{Toffoli}
\end{center}
\end{figure*}

Toffoli gate flips the state of the target qubit if the two controlled qubits both are in $|1\rangle$, and has no effect otherwise.
Figure \ref{Toffoli} depicts a scheme for implementing a Toffoli gate with an average success probability of 1/30 in the linear optical system.

Suppose two controlled photons, one target photon, and two auxiliary photons are initially prepared in the following states
\begin{eqnarray}              \label{eq11}
\begin{split}
|\psi\rangle_{c_{1in}}=\alpha_1|H\rangle_{c_{1in}}+\beta_1|V\rangle_{c_{1in}},
\end{split}
\end{eqnarray}
\begin{eqnarray}              \label{eq12}
\begin{split}
|\psi\rangle_{c_{2in}}=\alpha_2|H\rangle_{c_{2in}}+\beta_2|V\rangle_{c_{2in}},
\end{split}
\end{eqnarray}
\begin{eqnarray}              \label{eq13}
\begin{split}
|\psi\rangle_{t_{in}}=\alpha_3|H\rangle_{t_{in}}+\beta_3|V\rangle_{t_{in}},
\end{split}
\end{eqnarray}
\begin{eqnarray}              \label{eq14}
\begin{split}
|\psi\rangle_{a_1}=\frac{1}{\sqrt{2}}(|H\rangle_{a_1}+|V\rangle_{a_1}),
\end{split}
\end{eqnarray}
\begin{eqnarray}              \label{eq15}
\begin{split}
|\psi\rangle_{a_2}=\frac{1}{\sqrt{2}}(|H\rangle_{a_2}+|V\rangle_{a_2}),
\end{split}
\end{eqnarray}
where $|\alpha_1|^2+|\beta_1|^2=1$, $|\alpha_2|^2+|\beta_2|^2=1$, and $|\alpha_3|^2+|\beta_3|^2=1$.

In the first step, we employ PBS$_1$ to complete the parity-check measurement on the first controlled photon (emitted from spatial $c_{1in}$) and the first additional photon (emitted from spatial $a_1$), and choose the instance in which each outing mode contains exactly one photon. And then, PBS$_1$ converts the whole system from the initial state $|\Psi_0\rangle =|\psi\rangle_{c_{1in}}\otimes|\psi\rangle_{a_1}\otimes|\psi\rangle_{c_{2in}}\otimes|\psi\rangle_{a_2}\otimes|\psi\rangle_{t_{in}}$ to
\begin{eqnarray}              \label{eq16}
\begin{split}
|\Psi_1\rangle =&\frac{1}{\sqrt{2}} \big(\alpha_1|H\rangle_1|H\rangle_2+\beta_1|V\rangle_1|V\rangle_2 \big)\otimes\big(\alpha_2|H\rangle_{c_{2in}}+\beta_2|V\rangle_{c_{2in}}\big)\otimes\big(|H\rangle_{a_2}+|V\rangle_{a_2}\big) \\
               &\otimes\big(\alpha_3|H\rangle_{t_{in}}+\beta_3|V\rangle_{t_{in}}\big),
\end{split}
\end{eqnarray}
with a probability of 1/2.

In the second step, PBS$_2$ transmits $H_{c_{2in}}$-polarized component to PBS$_8$ and reflects $V_{c_{2in}}$-polarized component to spatial mode 3 for mixing with the components emitted from spatial mode 2 at PBS$_3$. After the photons emitted from spatial modes 2 and 3 experience the block composed of four HWP$^{22.5^\circ}$s and PBS$_3$, we choose the instance in which each of spatial modes 5 and 6 contains exactly one photon, and then the system will become  the normalization state
\begin{eqnarray}              \label{eq17}
\begin{split}
|\Psi_2\rangle =&\frac{1}{2\sqrt{2}} \big[ \big(\alpha_1\alpha_2|H\rangle_1|H\rangle_4|H\rangle_6+\beta_1\alpha_2|V\rangle_1|H\rangle_4|H\rangle_6 +\alpha_1\alpha_2|H\rangle_1|H\rangle_4|V\rangle_6 \\&
+\beta_1\alpha_2|V\rangle_1|H\rangle_4|V\rangle_6+\alpha_1\alpha_2|H\rangle_1|H\rangle_4|H\rangle_5-\beta_1\alpha_2|V\rangle_1|H\rangle_4|H\rangle_5 \\&-\alpha_1\alpha_2|H\rangle_1|H\rangle_4|V\rangle_5+\beta_1\alpha_2|V\rangle_1|H\rangle_4|V\rangle_5\big)
+\frac{1}{2} \big(\alpha_1\beta_2|H\rangle_1|V\rangle_5|H\rangle_6 \\
&+\beta_1\beta_2|V\rangle_1|H\rangle_5|H\rangle_6
+\alpha_1\beta_2|H\rangle_1|H\rangle_5|V\rangle_6+\beta_1\beta_2|V\rangle_1|V\rangle_5|V\rangle_6 \big)\big]\\
&\otimes\big(|H\rangle_{a_2}+|V\rangle_{a_2}\big)\otimes\big(\alpha_3|H\rangle_{t_{in}}+\beta_3|V\rangle_{t_{in}}\big),
\end{split}
\end{eqnarray}
with a probability of $\frac{1}{2}\times\frac{1+\alpha_2^2}{2}$.
In order to complete the Toffoli gate with unity fidelity in principle, we next reduce the amplitude of the photon emitted from spatial mode 5 to half by using a 50:50 beam splitter (BS). The unitary transformations of the BS can be described as
\begin{eqnarray}                  \label{eq18}
\begin{split}
&|H\rangle_5\xrightarrow{\text{BS}}\frac{1}{\sqrt{2}}\big(|H\rangle_{5'}+|H\rangle_{5''}\big), \quad
&|V\rangle_5\xrightarrow{\text{BS}}\frac{1}{\sqrt{2}}\big(|V\rangle_{5'}+|V\rangle_{5''}\big).
\end{split}
\end{eqnarray}
That is, BS yields the state
\begin{eqnarray}              \label{eq19}
\begin{split}
|\Psi_3\rangle =&\frac{1}{2\sqrt{2}} \big(\alpha_1\alpha_2|H\rangle_1|H\rangle_4+\beta_1\alpha_2|V\rangle_1|H\rangle_4 +\alpha_1\beta_2|H\rangle_1|V\rangle_{5'}\\&+\beta_1\beta_2|V\rangle_1|H\rangle_{5'}+\alpha_1\beta_2|H\rangle_1|V\rangle_{5''}+\beta_1\beta_2|V\rangle_1|H\rangle_{5''}\big)
\\&\otimes|H\rangle_6\otimes\big(|H\rangle_{a_2}+|V\rangle_{a_2}\big) \otimes\big(\alpha_3|H\rangle_{t_{in}}+\beta_3|V\rangle_{t_{in}}\big)\\
&+\frac{1}{2\sqrt{2}}\big(\alpha_1\alpha_2|H\rangle_1|H\rangle_4+\beta_1\alpha_2|V\rangle_1|H\rangle_4 +\alpha_1\beta_2|H\rangle_1|H\rangle_{5'}\\
&+\beta_1\beta_2|V\rangle_1|V\rangle_{5'}+\alpha_1\beta_2|H\rangle_1|H\rangle_{5''}+\beta_1\beta_2|V\rangle_1|V\rangle_{5''} \big)\\
&\otimes|V\rangle_6\otimes\big(|H\rangle_{a_2}+|V\rangle_{a_2}\big)\otimes \big(\alpha_3|H\rangle_{t_{in}}+\beta_3|V\rangle_{t_{in}}\big) \\
&+\frac{1}{4} \big(\alpha_1\alpha_2|H\rangle_1|H\rangle_4|H\rangle_{5'}-\alpha_1\alpha_2|H\rangle_1|H\rangle_4|V\rangle_{5'}-\beta_1\alpha_2|V\rangle_1|H\rangle_4|H\rangle_{5'}\\
&+\beta_1\alpha_2|V\rangle_1|H\rangle_4|V\rangle_{5'}+\alpha_1\alpha_2|H\rangle_1|H\rangle_4|H\rangle_{5''}-\alpha_1\alpha_2|H\rangle_1|H\rangle_4|V\rangle_{5''}\\&-\beta_1\alpha_2|V\rangle_1|H\rangle_4|H\rangle_{5''}
+\beta_1\alpha_2|V\rangle_1|H\rangle_4|V\rangle_{5''}\big)\\&\otimes\big(|H\rangle_{a_2}+|V\rangle_{a_2}\big) \otimes\big(\alpha_3|H\rangle_{t_{in}}+\beta_3|V\rangle_{t_{in}}\big).
\end{split}
\end{eqnarray}
If $D_{H_6}$  is triggered, the photon emitted from spatial mode $5'$  is led to PBS$_5$ to mix with the photon emitted from mode $a_2$, Eq. (\ref{eq19}) will collapse into the state
\begin{eqnarray}              \label{eq20}
\begin{split}
|\Psi_4\rangle =&\frac{1}{\sqrt{2}} \big(\alpha_1\alpha_2|H\rangle_1|H\rangle_4+\alpha_1\beta_2|H\rangle_1|V\rangle_{5'}+\beta_1\alpha_2|V\rangle_1|H\rangle_4+\beta_1\beta_2|V\rangle_1|H\rangle_{5'} \big)\\
&\otimes\big(|H\rangle_{a_2}+|V\rangle_{a_2}\big) \otimes\big(\alpha_3|H\rangle_{t_{in}}+\beta_3|V\rangle_{t_{in}}\big), \\
\end{split}
\end{eqnarray}
with a probability of $\frac{1}{2}\times\frac{1+\alpha_2^2}{2}\times \frac{1}{4}$. If $D_{V_6}$ is triggered, the photon emitted from spatial mode $5'$ will be applied a   feedback $\sigma_x$ operation to convert the state
\begin{eqnarray}              \label{eq20.5}
\begin{split}
|\Psi'_4\rangle =&\frac{1}{\sqrt{2}} \big(\alpha_1\alpha_2|H\rangle_1|H\rangle_4+\alpha_1\beta_2|H\rangle_1|H\rangle_{5'} +\beta_1\alpha_2|V\rangle_1|H\rangle_4+\beta_1\beta_2|V\rangle_1|V\rangle_{5'} \big)\\
&\otimes\big(|H\rangle_{a_2}+|V\rangle_{a_2}\big) \otimes\big(\alpha_3|H\rangle_{t_{in}}+\beta_3|V\rangle_{t_{in}}\big), \\
\end{split}
\end{eqnarray}
with a probability of $\frac{1}{2}\times\frac{1+\alpha_2^2}{2}\times \frac{1}{4}$ to Eq. (\ref{eq20}). The $\sigma_x$ operation can be achieved easily by an HWP$^{45^\circ}$ setting in mode $5'$.


In the third step, after PBS$_5$ completes the parity-check measurement on the photons emitted from spatial modes $5'$ and $a_2$, the instance in which the spatial mode 8 involves exactly one photon is chosen, and then the system will be in the following normalization state
\begin{eqnarray}              \label{eq21}
\begin{split}
|\Psi_5\rangle =& \big(\alpha_1\alpha_2|H\rangle_1|H\rangle_4|V\rangle_8+\alpha_1\beta_2|H\rangle_1|V\rangle_7|V\rangle_8 +\beta_1\alpha_2|V\rangle_1|H\rangle_4|V\rangle_8\\
&+\beta_1\beta_2|V\rangle_1|H\rangle_7|H\rangle_8 \big) \otimes\big(\alpha_3|H\rangle_{t_{in}}+\beta_3|V\rangle_{t_{in}}\big), \\
\end{split}
\end{eqnarray}
with a probability of $\frac{1}{2}\times\frac{1+\alpha_2^2}{2}\times (\frac{1}{4}+\frac{1}{4})\times \frac{1}{2}$.

In the fourth step,  the photons emitted from spatial modes 7 and 4 pass through HWP$^{67.5^\circ}$ and HWP$^{22.5^\circ}$ respectively, and then are fed into PBS$_8$. Before and after the photons emitted from spatial modes 8 and $t_{in}$ mix at PBS$_6$, four HWP$^{22.5^\circ}$s  are performed on them. The even-parity is chosen for the polarized photons in modes 11 and 12 after PBS$_6$. Therefore, before $D_{H_{12}}$ or  $D_{V_{12}}$ is fired, these elements induce the outing photons in the state
\begin{eqnarray}              \label{eq22}
\begin{split}
|\Psi_6\rangle =&\frac{1}{2} \big(\gamma_1|H\rangle_1|H\rangle_9|H\rangle_{11}+\gamma_2|H\rangle_1|H\rangle_9|V\rangle_{11} +\gamma_3|H\rangle_1|V\rangle_9|H\rangle_{11}\\&+\gamma_4|H\rangle_1|V\rangle_9|V\rangle_{11}+\gamma_5|V\rangle_1|H\rangle_9|H\rangle_{11}+\gamma_6|V\rangle_1|H\rangle_9|V\rangle_{11} \\&+\gamma_7|V\rangle_1|V\rangle_9|V\rangle_{11}+\gamma_8|V\rangle_1|V\rangle_9|H\rangle_{11}\big) \otimes       |V\rangle_{12}  \\
&+\frac{1}{2} \big(\gamma_1|H\rangle_1|H\rangle_9|V\rangle_{11}+\gamma_2|H\rangle_1|H\rangle_9|H\rangle_{11} +\gamma_3|H\rangle_1|V\rangle_9|V\rangle_{11}\\
&+\gamma_4|H\rangle_1|V\rangle_9|H\rangle_{11}+\gamma_5|V\rangle_1|H\rangle_9|V\rangle_{11}+\gamma_6|V\rangle_1|H\rangle_9|H\rangle_{11}  \\&+\gamma_7|V\rangle_1|V\rangle_9|H\rangle_{11}+\gamma_8|V\rangle_1|V\rangle_9|V\rangle_{11}\big)  \otimes    |H\rangle_{12}  \\
%
&+\frac{1}{2} \big(\gamma_1|H\rangle_1|V\rangle_{10}|H\rangle_{11}+\gamma_2|H\rangle_1|V\rangle_{10}|V\rangle_{11} +\gamma_3|H\rangle_1|H\rangle_{10}|H\rangle_{11}\\
&+\gamma_4|H\rangle_1|H\rangle_{10}|V\rangle_{11}+\gamma_5|V\rangle_1|V\rangle_{10}|H\rangle_{11}+\gamma_6|V\rangle_1|V\rangle_{10}|V\rangle_{11}  \\
& -\gamma_7|V\rangle_1|H\rangle_{10}|V\rangle_{11}-\gamma_8|V\rangle_1|H\rangle_{10}|H\rangle_{11}\big)\otimes  |V\rangle_{12}  \\
&+\frac{1}{2} \big(\gamma_1|H\rangle_1|V\rangle_{10}|V\rangle_{11}+\gamma_2|H\rangle_1|V\rangle_{10}|H\rangle_{11} +\gamma_3|H\rangle_1|H\rangle_{10}|V\rangle_{11}\\
&+\gamma_4|H\rangle_1|H\rangle_{10}|H\rangle_{11}+\gamma_5|V\rangle_1|V\rangle_{10}|V\rangle_{11}+\gamma_6|V\rangle_1|V\rangle_{10}|H\rangle_{11} \\
&-\gamma_7|V\rangle_1|H\rangle_{10}|H\rangle_{11}-\gamma_8|V\rangle_1|H\rangle_{10}|V\rangle_{11}\big)  \otimes  |H\rangle_{12},  \\
\end{split}
\end{eqnarray}
with a probability of $\frac{1}{2}\times\frac{1+\alpha_2^2}{2}\times (\frac{1}{4}+\frac{1}{4})\times \frac{1}{2}\times \frac{1}{2}$. Here, for simplicity, the coefficients are written as $\gamma_1=\alpha_1\alpha_2\alpha_3$, $\gamma_2=\alpha_1\alpha_2\beta_3$, $\gamma_3=\alpha_1\beta_2\alpha_3$, $\gamma_4=\alpha_1\beta_2\beta_3$, $\gamma_5=\beta_1\alpha_2\alpha_3$, $\gamma_6=\beta_1\alpha_2\beta_3$, $\gamma_7=\beta_1\beta_2\alpha_3$, and $\gamma_8=\beta_1\beta_2\beta_3$.
Half-wave plate HWP$^{67.5^\circ}$ induces the transformations
\begin{eqnarray}                  \label{eq23}
\begin{split}
&|H\rangle\xrightarrow{\text{HWP}^{67.5^\circ}}\frac{1}{\sqrt{2}}\big(|V\rangle-|H\rangle\big), \quad
&|V\rangle\xrightarrow{\text{HWP}^{67.5^\circ}}\frac{1}{\sqrt{2}}\big(|H\rangle+|V\rangle\big).
\end{split}
\end{eqnarray}
Based on Eq. (\ref{eq22}), one can see that when $D_{V_{12}}$ is fired, the outing photons from the spatial modes $1$, $9$, and $11$ kept are in the state
\begin{eqnarray}              \label{eq24}
\begin{split}
|\Psi_7\rangle =&\gamma_1|H\rangle_1|H\rangle_9|H\rangle_{11}+\gamma_2|H\rangle_1|H\rangle_9|V\rangle_{11}  +\gamma_3|H\rangle_1|V\rangle_9|H\rangle_{11}+\gamma_4|H\rangle_1|V\rangle_9|V\rangle_{11}\\
&+\gamma_5|V\rangle_1|H\rangle_9|H\rangle_{11}+\gamma_6|V\rangle_1|H\rangle_9|V\rangle_{11}  +\gamma_7|V\rangle_1|V\rangle_9|V\rangle_{11}+\gamma_8|V\rangle_1|V\rangle_9|H\rangle_{11},
\end{split}
\end{eqnarray}
with a probability of $\frac{1}{2}\times\frac{1+\alpha_2^2}{2}\times (\frac{1}{4}+\frac{1}{4})\times \frac{1}{2}\times \frac{1}{2}\times \frac{1}{4}$. That is, the probabilistic Toffoli gate is completed.

\begin{table} [htb]
\centering \caption{Measurement outcomes and corresponding feed-forward operations in mode $5'$ or 11 for realizing a Toffoli gate.  $I_2$ is an identity operation and $\sigma_x$ is a Pauli $X$ operation, which can be realized by an HWP setting at $45^\circ$.}
\begin{tabular}{ccccccc}
\hline  \hline

\multicolumn {2}{c}{Measurement}   & \qquad\qquad  & \multicolumn {2}{c}{Feed-forward} & \qquad\qquad  & Achieved   \\ \cline{1-2}  \cline{4-5}

     Detector   &  Detector        & \quad &  mode $5'$                                &   mode 11  & \qquad\qquad &result  \\

\hline

$D_{H_6}$  &  $D_{V_{12}}$  & \quad &  $I_2$                                       &   $I_2$       &  & Toffoli  \\

$D_{H_6}$  &  $D_{H_{12}}$  & \quad &  $I_2$                                       &   $\sigma_x$  &  & Toffoli \\

$D_{V_6}$  &  $D_{V_{12}}$  & \quad &  $\sigma_x$                                  &   $I_2$       &  & Toffoli  \\

$D_{V_6}$  &  $D_{H_{12}}$  & \quad &  $\sigma_x$                                  &   $\sigma_x$   &  & Toffoli  \\
                             \hline  \hline
\end{tabular}\label{table1}
\end{table}

When $D_{H_{12}}$ is fired,  the outing photons from the spatial modes $1$, $9$, and $11$ kept are in the state
\begin{eqnarray}              \label{eq25}
\begin{split}
|\Psi_7'\rangle =&\gamma_1|H\rangle_1|H\rangle_9|V\rangle_{11}+\gamma_2|H\rangle_1|H\rangle_9|H\rangle_{11}+\gamma_3|H\rangle_1|V\rangle_9|V\rangle_{11}+\gamma_4|H\rangle_1|V\rangle_9|H\rangle_{11}\\
&+\gamma_5|V\rangle_1|H\rangle_9|V\rangle_{11}+\gamma_6|V\rangle_1|H\rangle_9|H\rangle_{11}  +\gamma_7|V\rangle_1|V\rangle_9|H\rangle_{11}+\gamma_8|V\rangle_1|V\rangle_9|V\rangle_{11},
\end{split}
\end{eqnarray}
with a probability of $\frac{1}{2}\times\frac{1+\alpha_2^2}{2}\times (\frac{1}{4}+\frac{1}{4})\times \frac{1}{2}\times \frac{1}{2}\times \frac{1}{4}$.  To complete the Toffoli gate, a feed-forward $\sigma_x$ operation is applied to  photon in spatial mode 11.  The outcomes of measurement and corresponding
feed-forward operations for completing Toffoli gate are summarized in Tab. \ref{table1}. When $D_{10}$ is fired,  it means that the scheme fails.

\begin{figure} 
\begin{center}
\includegraphics[width=7.2 cm,angle=0]{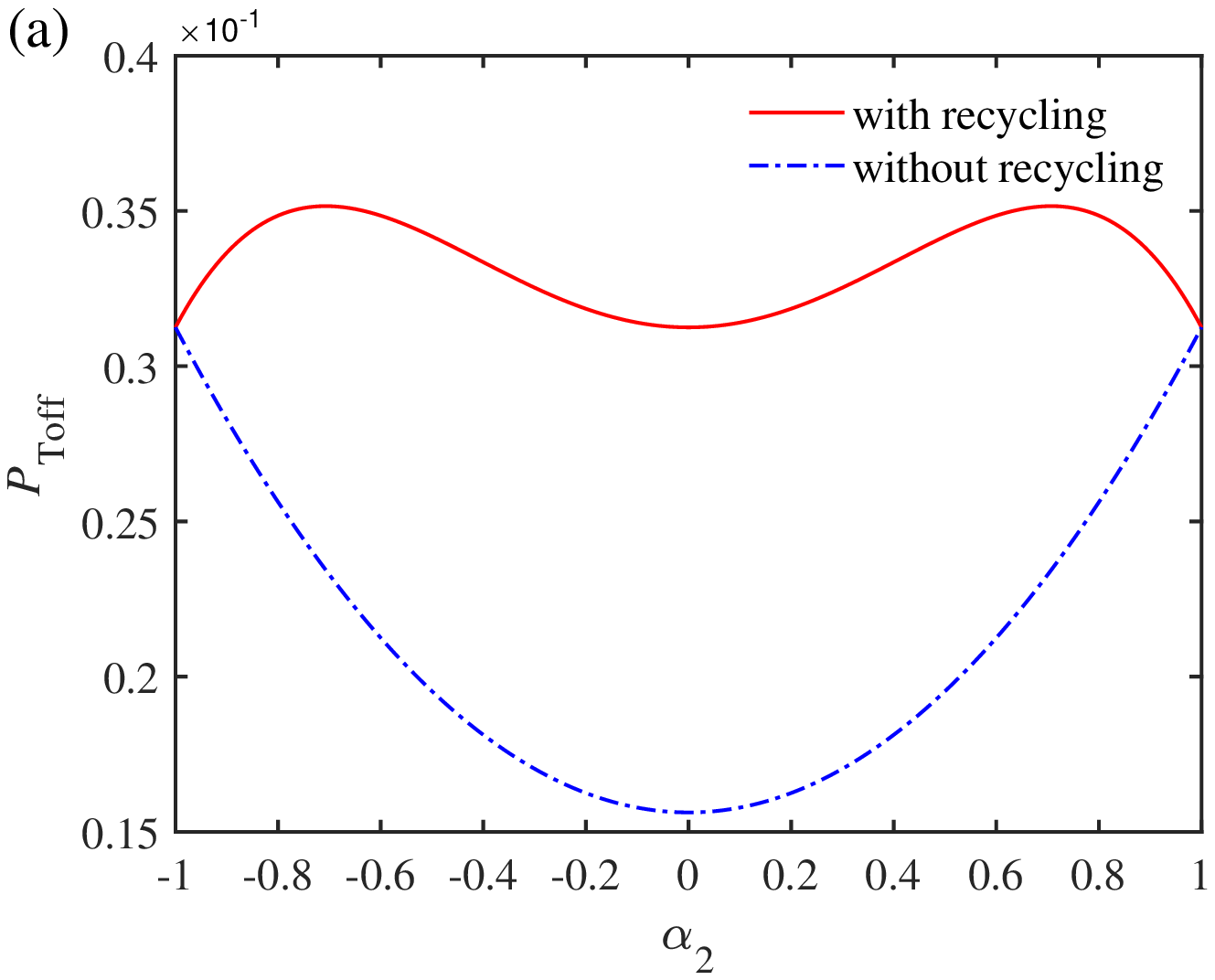}\quad
\includegraphics[width=7.2 cm,angle=0]{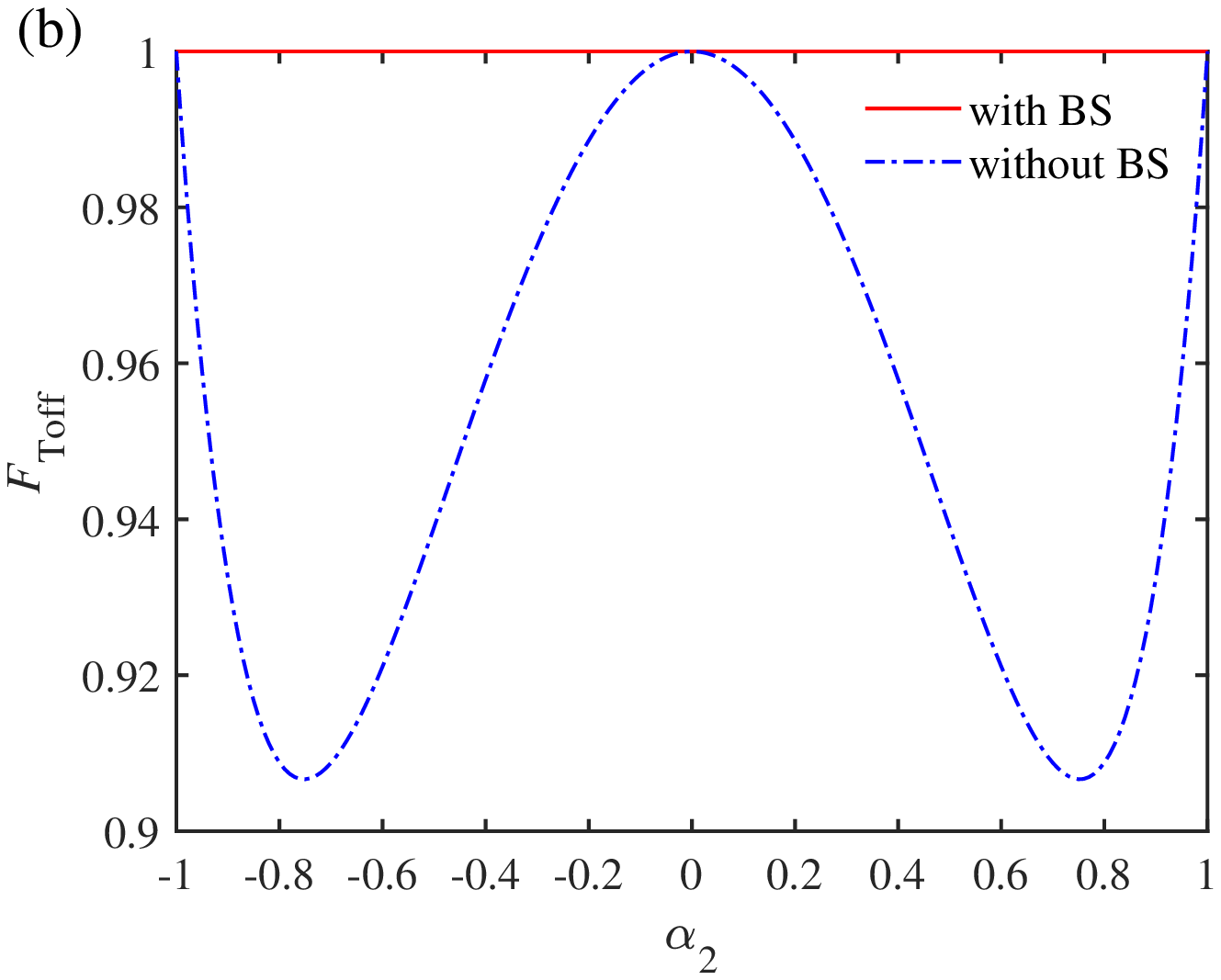}
\caption{(a) The relations between the success probability of Toffoli gate ($P_\text{Toff}$) and parameter $\alpha_2$ with/without recycling. The blue dashed curve indicates  $P_\text{Toff}=\frac{1+\alpha_2^2}{64}$ without recycling the photons in mode $5''$, while the red solid curve indicates  $P_\text{Toff}=\frac{(1+\alpha_2^2)(2-\alpha_2^2)}{64}$ with recycling the photons in mode $5''$. (b) The relations between the fidelity of Toffoli gate ($F_\text{Toff}$) and parameter $\alpha_2$ with/without BS. The blue dashed and red solid curves indicate the fidelity of the Toffoli gate without and with BS, respectively.} \label{PToff}
\end{center}
\end{figure}

We evaluate the performance of the scheme by characterizing success probability and fidelity of Toffoli gate, respectively. The success probability of Toffoli gate is defined as $P_\text{Toff}=n_\text{out}/n_\text{in}$, where $n_\text{out}$ and $n_\text{in}$ are the number of output photon and input photon, respectively. The fidelity of Toffoli gate is defined as $F_\text{Toff}=|\langle\varphi_\text{real}|\varphi_\text{ideal}\rangle|$, where $\varphi_\text{real}$ and $\varphi_\text{ideal}$ are the real normalized output state of the gate and ideal normalized output state of the gate, respectively.
The quantum circuit shown in Fig. \ref{Toffoli} completed a post-selection Toffoli gate with a success probability of $P_\text{Toff}=\frac{1}{2}\times\frac{1+\alpha_2^2}{2}\times\frac{1}{2}\times\frac{1}{2}\times\frac{1}{2}\times\frac{1}{2}= \frac{1+\alpha_2^2}{64}$ and unity fidelity in principle.
Alternately, the success probability of the gate with unity fidelity can be further enhanced to $\frac{1}{2}\times\frac{1+\alpha_2^2}{2}\times\frac{2-\alpha_2^2}{2}\times\frac{1}{2}\times\frac{1}{2}\times\frac{1}{2}= \frac{(1+\alpha_2^2)(2-\alpha_2^2)}{64}$ by recycling the photons emitted from spatial mode $5''$ into the PBS$_5$  because the state has the same form as the state shown in Eq. (\ref{eq20}). From Fig. \ref{PToff}(a),  one can see that the minimum $P_\text{Toff}$ without recycling is approximately $\frac{1}{64}$, the maximum $P_\text{Toff}$ without recycling reaches approximately $\frac{1}{32}$, and the average $P_\text{Toff}$ without recycling is $\frac{1}{2} \int_{-1}^{1}\frac{1+\alpha_2^2}{64}d\alpha_2=\frac{1}{48}$. In contrast, with recycling one, the minimum $P_\text{Toff}$  is approximately $\frac{1}{32}$,  the maximum $P_\text{Toff}$ can reach $\frac{9}{256}$, and the average $P_\text{Toff}$ is $\frac{1}{30}$. The success probability of our Toffoli gate is much higher than previous works \cite{T-Fiurasek,T-PRA,T-NatPhy,T-Liu,li2022quantum}.
As shown in Fig. \ref{PToff}(b), if the BS is discarded in Fig. \ref{Toffoli}, the fidelity of Toffoli gate will reduce from unity (red solid curve) to $F_\text{Toff}=1-\alpha_2^2+\frac{\alpha_2^2}{\sqrt{2-\alpha_2^2}}$ (blue dashed curve), and the success probability will be $P_\text{Toff}=\frac{(1+\alpha_2^2)(2-\alpha_2^2)}{64}$, which is higher than the without recycling one and equals to the with recycling one.
From Fig. \ref{PToff}(b), one can see that the fidelity of the Toffoli gate keeps unity for the scheme with  BS in principle, while the average fidelity of the Toffoli gate over $\alpha_2$ is $\frac{3\pi+2}{12}\approx95.2\%$ for the scheme without BS.

\begin{table}[htb]
\centering\caption{A comparison of proposed post-selected  CPF  gate with linear optics and previous schemes.}
\begin{tabular}{lcccccc}

\hline  \hline
Proposed    &        & Ancillary    &        &Success        &           & Achieved         \\
schemes     &        & photons      &        &probability   &           & results           \\

\hline

  Refs. \cite{CNOT-Bell-1/4-1,CNOT-Bell-1/4-2,CNOT-Bell-1/4-3,CNOT-Bell-1/4-4,CNOT-Bell-1/4-5}    &   & A Bell-state       &   &  1/4   &   &  CNOT  \\
  Refs. \cite{CNOT-1/9-2,CNOT-1/9-4,CNOT-1/9-5}    &   & No   &   & 1/9      &    &  CPF   \\
  Refs. \cite{CNOT-1/9-3,CNOT-1/9-6,CNOT-1/9-7}    &   & No   &   & 1/9       &    &  CNOT   \\
  Refs. \cite{CNOT-1/8-1,CNOT-1/8-2}    &   & Two single photons   &  & 1/8       &    &  CNOT   \\
  Ref. \cite{CNOT-1/16}   &   & Two single photons                                                &    &  1/16     &       &   CNOT  \\
  Ref. \cite{T-Liu} &   & No    &   & 1/8       &    &  CNOT   \\
  Ref. \cite{li2022quantum}  &   &  No     &   &   1/9       &    &  CNOT  \\
  This work   &   & A single photon   &    &  1/4    &       &  CPF  \\

\hline  \hline
\end{tabular}\label{compareCPF}
\end{table}

\begin{table}[htb]
\centering\caption{A comparison of proposed post-selected  Toffoli  gate with linear optics and previous schemes.}
\begin{tabular}{lcccccc}

\hline  \hline
Proposed        &  \quad\;  & Ancillary        & \quad\;   &Success             \\
schemes          & \quad\;  & photons          & \quad\;   &probability         \\

\hline

  Fiur\'{a}\v{s}ek \cite{T-Fiurasek}       &     & No                     &      &  1/133                \\
  Ralph \emph{et al.} \cite{T-PRA}         &     & No                     &      &  1/72                  \\
  Ralph \emph{et al.} \cite{T-PRA}         &     & Two Bell-states         &      &  1/32                  \\
  Lanyon \emph{et al.} \cite{T-NatPhy}     &     & No                     &      &  1/72                  \\
  Liu \emph{et al.} \cite{T-Liu}           &     & No                     &      &  1/64                   \\
  Li \emph{et al.} \cite{li2022quantum}    &     &  No                      &      &  1/60                   \\
  This work                                &     & Two single photons     &      &  1/30                  \\

\hline  \hline
\end{tabular}\label{compareToffoli}
\end{table}

\section{Conclusion} \label{sec4}

We have proposed two schemes to implement post-selected CPF and Toffoli gates in the coincidence basis by solely using linear optics. The comparisons between our proposed CPF gate and Toffoli gate and previous schemes are presented in Tab. \ref{compareCPF} and Tab. \ref{compareToffoli}, respectively. A  maximally entangled photon pair \cite{CNOT-Bell-1/4-1,CNOT-Bell-1/4-2,CNOT-Bell-1/4-3,CNOT-Bell-1/4-4,CNOT-Bell-1/4-5} (or two single photons \cite{CNOT-1/8-1,CNOT-1/8-2}) is necessary for implementing a CNOT gate with the current existing maximum success probability of 1/4 (or 1/8).
Only one auxiliary single photon is introduced to accomplish our CPF gate with the success probability of 1/4. In addition, our approach to implement Toffoli gate is much more efficient than the synthesis one. Assisted by two independent single photons, our Toffoli gate is constructed  with the current maximum success probability of 1/30.

We note that the success condition of the gates can only be verified by the final detection of outing photons using the photon-number-resolving detector (PNRD) \cite{fitch2003photon,divochiy2008superconducting,nehra2020photon}. This coincidence basis operation without quantum nondemolition detection causes an unscalable approach to the construction of quantum gate, but it provides a convenient means of experimental test of  optical CPF gate and Toffoli gate. The PNRD can distinguish the number of photons, which is an important resource for many quantum information processing tasks realized in experiments \cite{CNOT-Bell-1/4-5,CNOT-1/8-1,CNOT-1/8-2,pan2001entanglement}. Our presented two architectures open an alternative insight into probabilistic quantum gates using linear optical elements and suggest that they maybe have various applications in photonic quantum information processing.

\section*{Acknowledgments}

This work was supported by the National Natural Science Foundation of China under Grant No. 11604012, the Fundamental Research Funds for the Central Universities under Grants FRF-TP-19-011A3.

\bibliographystyle{SciPost_bibstyle}
\bibliography{mybibliography}

\end{document}